\documentclass[%
 aip,
 jmp,%
 amsmath,amssymb,
 reprint,%
]{revtex4-1}

\usepackage{graphicx}
\usepackage{xcolor}
\usepackage{url}
\usepackage{color}
\usepackage{dcolumn}
\usepackage{bm}
\usepackage[version=3]{mhchem}
\usepackage[left=1.5cm, right=1.5cm, top=1.785cm, bottom=2.0cm]{geometry}
\usepackage{balance}
\usepackage[capitalise]{cleveref} 
\usepackage{times,mathptmx}
\usepackage{graphicx} 
\usepackage{epstopdf}
\usepackage[euler]{textgreek}
\usepackage{lastpage}
\newcommand*{\myfont}{\fontfamily{phv}\selectfont}
\DeclareTextFontCommand{\textmyfont}{\myfont}
\newcommand{\q}[1]{``#1''}
\DeclareMathOperator*{\argmin}{argmin}
\newcommand{\rachomega}[1]{\textmyfont{\textomega}$_#1$}
\newcommand{\rachalpha}[1]{\textmyfont{\textalpha}$_#1$}
\newcommand{\rachphi}{\textmyfont{\straightphi}}
\newcommand{\rachpsi}{\textmyfont{\textpsi}}
\newcommand{\myfontsp}[2]{\ensuremath{\textmyfont{#1}_{\textmyfont{#2}}}}
\newcommand{\myatm}[2]{\ensuremath{\ce{#1}_{#2}}}
\newcommand{\mysubfigref}[2]{\cref{#1}-#2}

\begin{document}

\preprint{AIP/123-QED}

\title{Mapping and Classifying Molecules from \\a High-Throughput Structural Database}

\author{Sandip De}
\email{sandip.de@epfl.ch}
 \affiliation{National Center
for Computational Design and Discovery of Novel Materials (MARVEL)}
 \affiliation{Laboratory of Computational Science and Modelling, Institute of Materials, Ecole Polytechnique F\'ed\'erale de Lausanne, Lausanne, Switzerland}
 \author{Felix Musil}
 \affiliation{Laboratory of Computational Science and Modelling, Institute of Materials, Ecole Polytechnique F\'ed\'erale de Lausanne, Lausanne, Switzerland}
\author{Teresa Ingram}%
 \affiliation{Theory Department of the
Fritz Haber Institute, Faradayweg 4-6
D-14195 Berlin-Dahlem, Germany}

\author{Carsten Baldauf}
\affiliation{Theory Department of the
Fritz Haber Institute, Faradayweg 4-6
D-14195 Berlin-Dahlem, Germany}

\author{Michele Ceriotti}
 \affiliation{National Center
for Computational Design and Discovery of Novel Materials (MARVEL)}
 \affiliation{Laboratory of Computational Science and Modelling, Institute of Materials, Ecole Polytechnique F\'ed\'erale de Lausanne, Lausanne, Switzerland}

\date{\today}

\begin{abstract}
High-throughput computational materials design
promises to greatly accelerate the process of  
discovering new materials and compounds,  
and of optimizing their properties. The large 
databases of structures and properties that result from
computational searches, as well as the agglomeration of 
data of heterogeneous provenance leads to considerable 
challenges when it comes to navigating the database, 
representing its structure at a glance, understanding 
structure-property relations, eliminating duplicates and 
identifying inconsistencies. Here we present a case study, 
based on a data set of conformers of amino acids and dipeptides, of how 
machine-learning techniques can help addressing these 
issues. We will exploit a recently-developed strategy to 
define a metric between structures, and use it as the basis 
of both clustering and dimensionality reduction techniques 
-- showing how these can help reveal structure-property 
relations, identify outliers and inconsistent structures, 
and rationalise how perturbations (e.g. binding of ions to 
the molecule) affect the stability of different conformers.
\end{abstract}

\pacs{Valid PACS appear here}
\maketitle
    
Computational materials design promises to greatly accelerate the 
discovery of materials and molecules with novel, optimized or custom-tailored properties. 
With this goal in mind, several community efforts have emerged
over the past few years \cite{aiida,clean_en,esp,oqmd,pauling,matgenome1,matgenome2,PhysRevLett.108.058301} that aim at generating, and/or storing
large amounts of simulation data in publicly available databases\cite{PhysRevLett.114.105503,
PhysRevB.92.014106,PhysRevB.92.094306,kusne15screport,
ramkrsinan_2014sd,PhysRevB.90.155136,PhysRevB.90.155136,db-paper}. 
The development of these repositories of structural data, and of 
associated materials properties (e.g. formation energy, band gap,
polarizability, \ldots) poses considerable challenges, from 
the points of view of guaranteeing consistency, accuracy and 
reliability of the stored information, as well as that of
extracting intuitive insight onto the behavior of a given class
of materials and of data-mining  in search of 
compounds that exhibit the desired properties
or that are somehow interesting or unexpected. 

In order to automate these tasks -- which is
necessary to unlock the full potential of 
computational materials databases that
can easily contain millions of distinct structures
-- a number of different machine-learning algorithms
have been developed, or adapted to the specific
requirements of this field~\cite{rodr-laio14science,
clustering-rui,gang_clustering,cartography,
prasanna-sd,ferg+10pnas,ceri+11pnas,trib+12pnas,
ceri+13jctc,rohr+arpc13}.
A fundamental ingredient in all of these approaches
is a concise mathematical representation of a 
molecular or crystalline structure, that can take 
the form of fingerprints (low-dimensional representation
of the structure of the atoms) or more abstract 
measures of the (dis-)similarity between elements
in the database, such as distance or kernel functions. 

In the present manuscript we will present a demonstration
of how a combination of a very general approach to 
quantify structural dissimilarity~\cite{de+16pccp} can
be combined with non-linear dimensionality
reduction and clustering techniques to address
the challenges of navigating a database of
molecular conformers,
checking its internal consistency and
rationalising structure-property relations. 
Even though we will focus in particular on 
a energy/structure data set of amino acid and dipeptide conformers obtained by
an \textit{ab initio} structure search~\cite{aminodb,db-paper},
many of the observations we will infer are general,
and provide insight on the application of 
machine-learning techniques to the analysis 
of molecular and materials databases generated 
by high-throughput computations.

\section{A Toolbox for Database Analysis}
 
In this section we will describe a specific 
combination of atomic structure representations and
of analysis techniques that can be used to deal
effectively with large databases of materials and
molecules. We will however also briefly 
summarize some of the alternative approaches that
could be used to substitute different components
of our tool chain. 
  
\subsection{Fingerprints and Structural Dissimilarity}
  
 The most crucial and basic element in any 
 structural analysis algorithm is to introduce a 
 metric to measure (dis)similarity between two atomic 
 configurations. Many options are available, with 
 different levels of complexity and generality, 
 starting from the commonly used Root Mean Square (RMS) 
 distance. In order to deal with symmetry operations
 or condensed phase structures, several ``fingerprint''
 frameworks have been developed \cite{sprint,PhysRevLett.108.058301,PhysRevB.90.104108,PhysRevB.89.235411,pilania13screport,PhysRevB.88.054104,rupp+07jcim,hirn+15arxiv,qm7b,PhysRevLett.108.253002,PhysRevB.92.045131,anatoleIJQC,bag.of.bonds,newstefan},
 that assign a unique vector of order parameters to each
 molecular or crystalline configuration: a metric can 
 then be easily built by taking some norm of the 
 difference between fingerprint vectors. 
 Any of these distances could be taken as the basis
 of the classification and mapping algorithms that
 we will describe in what follows. 
 
 In this paper we will use instead a very 
 flexible framework
 (REMatch-SOAP)  that is based on the definition of 
 an environment similarity matrix $C_{ij}(A,B)$, 
 which contains the
 complete information on the pair-wise similarity of 
 the environment of each of the atoms within the 
 molecules $A$ and $B$. In our framework, the 
 similarity between two local environments $\mathcal{X}^A_{i}$ and $\mathcal{X}^B_{j}$ 
 is computed  using the SOAP kernel~\cite{PhysRevB.90.104108}
 \begin{equation}
   C_{ij}(A,B)=k(\mathcal{X}^A_{i},\mathcal{X}^B_{j}).
 \end{equation}
The REMatch kernel is then defined as the 
following weighted combination of the elements
of $\mathbf{C}(A,B)$
  \begin{equation}
\begin{split}
   &\hat{K}^{\gamma}(A,B) = \operatorname{Tr}\mathbf{P}^\gamma\mathbf{C}(A,B), \\
   &\mathbf{P}^\gamma =\operatorname*{argmin}_{\mathbf{P}\in \mathcal{U}(N,N)} \sum_{ij}   P_{ij} \left(1-C_{ij}+\gamma \ln P_{ij}\right), \\
   & C_{ij}(A,B)=k\left(\mathcal{X}^A_{i},\mathcal{X}^B_{j}\right),
   \end{split}
   \label{eq:k-regmatch}
\end{equation}
where the optimal combination is obtained by 
searching over the doubly stochastic matrices
$\mathcal{U}(N,N)$ the one that minimizes the 
discrepancy between matching pairs of environments
subject to a regularization based on the matrix
information-entropy 
$E(\mathbf{P})=-\sum_{ij}P_{ij}\ln P_{ij}$~\cite{cutu13nips}. 
Once a kernel between two configurations has been
defined, it is then possible to introduce 
a kernel distance
\begin{equation}
    D(A,B)=\sqrt{\hat{K}^{\gamma}(A,A)+\hat{K}^{\gamma}(B,B)-2\hat{K}^{\gamma}(A,B)},
    \label{eq:dissim}
\end{equation}
that we will use as the metric for representing
and clustering structures from a database.

As discussed in Ref.~\cite{de+16pccp}, 
the choice of a SOAP kernel as the definition of
an environment similarity provides at the same time
great generality -- it can be seamlessly applied
to both molecules and solids -- and elbowroom for 
fine-tuning -- ranging from setting an appropriate 
cutoff distance to circumscribe an
environment to the introduction of an alchemical
similarity kernel that translates the notion that
different chemical species can behave similarly
with respect to the properties of interest.



\subsection{Mapping the Structural Landscape of a Database}
The dissimilarity between the $N$ atomic configurations in a 
database contains a large amount of information on the structural
relations between the database items. However, this information 
is not readily interpretable, as it is encoded as a $N^2$ matrix 
of numbers. Several methods are available to process dissimilarity 
information into a form that can be understood more intuitively. 
A first approach involves building a low-dimensional ``map'', 
where each point corresponds to one of the structures in the database 
and where the (Euclidean) distances between points represents the 
information on the pairwise dissimilarity matrix. 

Several methods have been proposed over the years to 
solve this dimensionality reduction problem, starting from
principal component analysis \cite{pca-WOLD198737} and the equivalent
linear multi-dimensional scaling~\cite{Kruskal1964}, and proceeding
to non-linear generalizations of the idea, such as 
ISOMAP~\cite{tene+00science}, diffusion maps~\cite{coif+05pnas},
kernel PCA~\cite{scho+98nc}. In this manuscript, we will
use sketchmap~\cite{ceri+11pnas,trib+12pnas,ceri+13jctc}, 
a method in which one iteratively optimizes the
objective function 
\begin{equation}
S^2 = \sum_{ij} \left[F\left[D(X_i,X_j)\right]-
f\left[d(x_i,x_j)\right]\right]^2, \label{eq:smap}
\end{equation}
that measures the mismatch of the dissimilarity
between atomic configurations $D(X_i,X_j)$ with the 
dissimilarity (typically just the Euclidean distance)
between the corresponding low-dimensional projections 
 $\left\{x_i\right\}$. 
The procedure is very similar to multi-dimensional 
scaling, except for the appearance of the transformations
$F$ and $f$, which are non-linear sigmoid functions 
of the form:
\begin{equation}
F(r) = 1 - ( 1 + (2^{a/b} - 1)(r/\sigma)^a )^{-b/a}.
\end{equation}
The non-linear transformation focuses the optimization 
of \cref{eq:smap} on the most significant distances 
(typically those of the order of $\sigma$), and 
disregards local distortions (e.g. induced by thermal
fluctuations or by incomplete convergence of a 
geometry optimization) and the relation between
completely unrelated portions of configuration landscape. 
The maps that we report in this work will be labeled synthetically 
using the notation {\texttt{$\sigma$-A\_B-a\_b}},
 where $A$ and $B$ denote the exponents used for the
 high-dimensional function $F$, $a$ and $b$ denote the 
 exponents for the low-dimensional function $f$, and 
 $\sigma$ the threshold for the switching function.
The choice of these parameters of the sigmoid functions 
are discussed in detail elsewhere~\cite{ceri+13jctc}. 
In practice $A$, $B$, $a$ and $b$ have relatively small
effect on the projection, and can be optimized and kept
fixed for systems belonging to the same family. 
Since the structures we consider here are 
minimum-energy configurations, and there are no thermal
fluctuations that should be filtered out, we set
$A=a=1$ (so that at short range the algorithm will
still try to represent distances faithfully) and
set the long-range exponents to $B=b=4$.
The parameter $\sigma$ is the one to which 
sketchmap is most sensitive, and 
needs to be tuned for each system separately. To 
automate the process of building sketchmaps of large 
amount of subsets of the database, we have used a
simple heuristic procedure for determining the value 
of $\sigma$ automatically. Following the prescriptions
in Ref.~\cite{ceri+13jctc}, we first compute the 
histogram of distances in the dissimilarity matrix
of each molecular set,
and detect the dissimilarity value ($D_{max}$)
corresponding to the peak value of the histogram.
We then set the value of $\sigma$ to $0.8D_{max}$.

\subsection{Hierarchical Clustering Representation}
As we will demonstrate below, sketchmap provides a
remarkably informative two-dimensional representation 
of structures in a data set, making it possible to 
identify groups of similar configurations, outliers,
as well as to investigate structure-property relations.
An alternative approach to navigate a set of structures
based on the dissimilarity matrix is to use clustering
algorithms, that identify groups of objects 
having similar properties to hint at the presence of
recurring motifs underlying the behavior of the 
system. 

A considerable number of clustering algorithms 
have been developed over the last few 
decades~\cite{Jain:1999:DCR:331499.331504,clustering-rui,aggarwal2013data}, 
including connectivity models \cite{WIDM:WIDM53} (i.e. hierarchical clustering), centroid models \cite{kmeans1,kmeans2,kmeans3} (i.e. k-means algorithm)
and density based models \cite{dbscan,optics,rodr-laio14science}.

Clustering models based on connectivity information 
such as hierarchical (or agglomerative) clustering 
\cite{WIDM:WIDM53} are particularly suited for this purpose 
and we will focus only on this type of clustering in this 
paper. Starting from each configurations as its own cluster, 
the hierarchical clustering algorithms iteratively 
aggregate clusters together based on some assessment of
their similarity. 
Similarity between two individual structures can be obviously
measured by their distance $D(A,B)$. The distance
between two \emph{clusters}, however, can be defined in many
different ways. In our study, we will use in particular 
the RMS dissimilarity between the pair of members of the 
two clusters. The linkage distance $\Delta$ between two clusters 
$\mathbb{X}=\left\{X_i\right\}$ and $\mathbb{Y}=\left\{Y_i\right\}$
is then defined as:
  \begin{equation}
      \Delta(\mathbb{X},\mathbb{Y})=
      \sqrt{ \frac{1}{N_\mathbb{X} N_\mathbb{Y}}\sum_{X \in \mathbb{X} , Y \in \mathbb{Y}}{D^2(X,Y)} },
      \label{eq:dendist}
  \end{equation}
where $N_\mathbb{X}$ and $N_\mathbb{Y}$ are the total 
number of configurations within each cluster.
$D(X,Y)$ is the dissimilarity between the two 
configurations, as computed e.g. from the REMatch-SOAP kernel. 
The complexity of this type of clustering  
($\mathcal{O}(N^2log(N))$) is relatively cheaper than 
dimensionality reduction algorithms like sketchmap 
($\mathcal{O}(N^3)$). Both procedures can be greatly
accelerated by selecting a subset of the configurations
(e.g. by farthest point sampling, with the possibility of weighting based on density information~\cite{ceri+13jctc}) which is then
subject to either dimensionality reduction or 
hierarchical clustering. 

 The results of a hierarchical clustering procedure
 can be represented in a \q{dendrogram} plot, that conveys
 visually the sequence of agglomerative clustering operations
 and the linkage distance at each step. 
 The x-axis represents the index of the structures, 
 sorted in a way so that at each stage the clusters being
 joint are adjacent. Each merge operation is 
 represented by a line joining the two underlying 
 clusters, with the $y$ position of the line representing
 the linkage distance for that pair, as defined by
 \cref{eq:dendist}. 
 In this kind of representation, at the bottom of the 
 dendrogram, each structure can be thought of as an individual 
 cluster containing only one item. 
 Clusters are then merged iteratively, selecting at 
 each step the pair of clusters that are closest to 
 each other. This  operation is repeated until all the
 clusters collapse into one single group that encompasses
 all the structures in the database, thus 
 completing the dendrogram. To avoid overcrowding the 
 bottom of the plot, one can hide the part 
 that corresponds to very small linkage distances, while
 still graphically visualising the size of the clusters
 by drawing bars that encompass the associated structures. 
  Since the ``leaves'' of this dendrogram correspond to 
 individual configurations, it is possible to complement
 the dendrogram with color-coded bar plots that represent
 the value of different properties of each structure, 
 thereby giving a clear picture of the relation between
 structural clustering and the different properties. 

 
In order to understand the basic motifs of a particular 
cluster $\mathbb{X}$, it is very useful to select one of 
its structures that is as representative as possible of 
the entire subset.
In case where stability estimates are
available, such structure may be the 
lowest-energy structure in the cluster.
For a definition that is based purely
on conformational or configurational information, 
the most representative structure 
$RS\left(\mathbb{X}\right)$ could be defined,
as the item having the minimum mean square dissimilarity 
with respect to all other members of $\mathbb{X}$, i.e.
 \begin{equation}
      RS\left(\mathbb{X}\right)= \argmin_{X_1\in \mathbb{X}}\left[\frac{1}{N_\mathbb{X}}\sum_{X_2\in \mathbb{X}}D^2(X_1,X_2) \right].
      \label{eq:rep-struc}
\end{equation}

Representative structures can be defined at 
each level of the hierarchy, and can therefore
be very useful in navigating the database, 
and understanding what are its
most crucial structural features. The spread 
of the cluster around $RS\left(\mathbb{X}\right)$,
 \begin{equation}
      \sigma_D\left(\mathbb{X}\right)= \sqrt{\frac{1}{N_\mathbb{X}}\sum_{X\in \mathbb{X}}D^2(X,RS(\mathbb{X})) },
      \label{eq:sigma-x}
\end{equation} 
 can be used to quantify the range of structural
 landscape that is covered by the 
 cluster. 
 
  Another important aspect of database analysis 
  is `outlier detection'~\cite{outlier1,outlier2,outlier3,outlier4,outlier5,outlier6}. 
  An ``outlier'' configuration is defined as a 
  configuration which is different from most of 
  the configurations in the database.
  Outlier configurations are very important as 
  they are likely to have unique structural motif in the whole database and are thus interesting 
  for structure prediction applications. They 
  also could  represent chemical changes or 
  indicate inconsistent configurations which are 
  likely to be ``errors'' in the database. 
  
  In the following sections, we will present
examples of how these different analyses can be
applied to different subsets of structures taken
from a database of amino acid and dipeptide conformers.
 
\section{Analysis of a database} 
This work is based on a first-principles derived 
structure/energy data set with conformers of 
twenty proteinogenic amino acids and dipeptides, 
as well as their interactions with a series of 
divalent cations\cite{aminodb} (Ca$^{2+}$, Ba$^{2+}$, Sr$^{2+}$, Cd$^{2+}$, Pb$^{2+}$, Hg$^{2+}$).
The potential-energy surfaces (PES) of 280 
systems were explored in a wide relative energy 
range of up to 4\,eV (390\,kJ/mol), summing up to 
an overall of 45,892 stationary points on the 
respective potential-energy 
surfaces~\cite{db-paper}.
The underlying energetics were calculated by 
applying density-functional theory (DFT) in the 
generalized gradient approximation corrected for 
long-range van der Waals interactions\cite{cpc180_2175,prl77_3865,prl102_73005} (PBE+vdW).
A number of theory-theory and theory-experiment 
comparisons have shown the applicability of the 
method to amino acid and peptide 
systems~\cite{prl106_118102,cej19_11224,db-paper,C4CP05541A,C4CP05216A,doi:10.1021/jp412055r,0953-8984-27-49-493002}.
Such dominantly manually curated data is a great 
starting point for studying materials and molecules across chemical 
space~\cite{ropo_scirep_2016}, yet automated techniques are needed to extract unbiased and hypothesis-free  trends and to check consistency of the underlying data.

\begin{figure}[!htbp]
 \centering
 \includegraphics[width=0.45\textwidth]{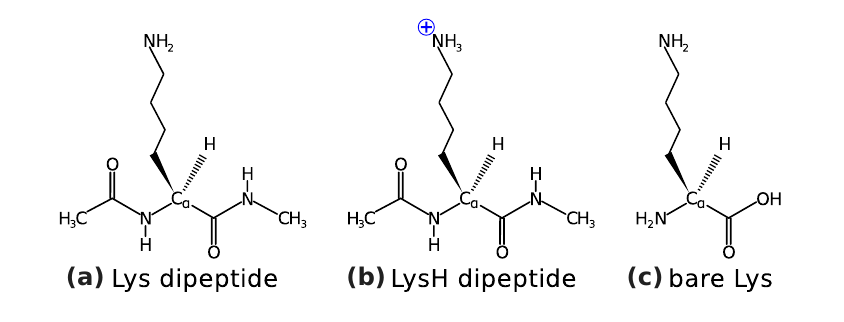}
    \caption{The lysine building block was studied in three forms: (a)~uncharged dipeptide, (b)~protonated dipeptide, and (c)~uncapped and uncharged amino acid.}
 \label{fig:lysines}
\end{figure}

In this study we focus on the amino acid 
lysine (in short Lys) and investigate basic 
structural motifs of three forms, see Fig.~\ref{fig:lysines}.
Furthermore, the machine learning techniques 
introduced in this work are used to detect 
the impact of perturbations (here Ca$^{2+}$ 
cations) on the structural properties of the 
unperturbed systems. Finally, we demonstrate 
how the approach can also be applied to discover 
inconsistencies and outliers in the database.
Hierarchical classifications and sketchmap 
projections for all the proteinogenic amino acids in the 
database are given in the Supporting Information.

\subsection{Finding the Dominant Features of a Structual Landscape}
 \label{sect:basic}
 
  \subsubsection{Lysine Dipeptide}
   \begin{figure*}[!htbp]
       \centering
       \includegraphics[width=\textwidth]{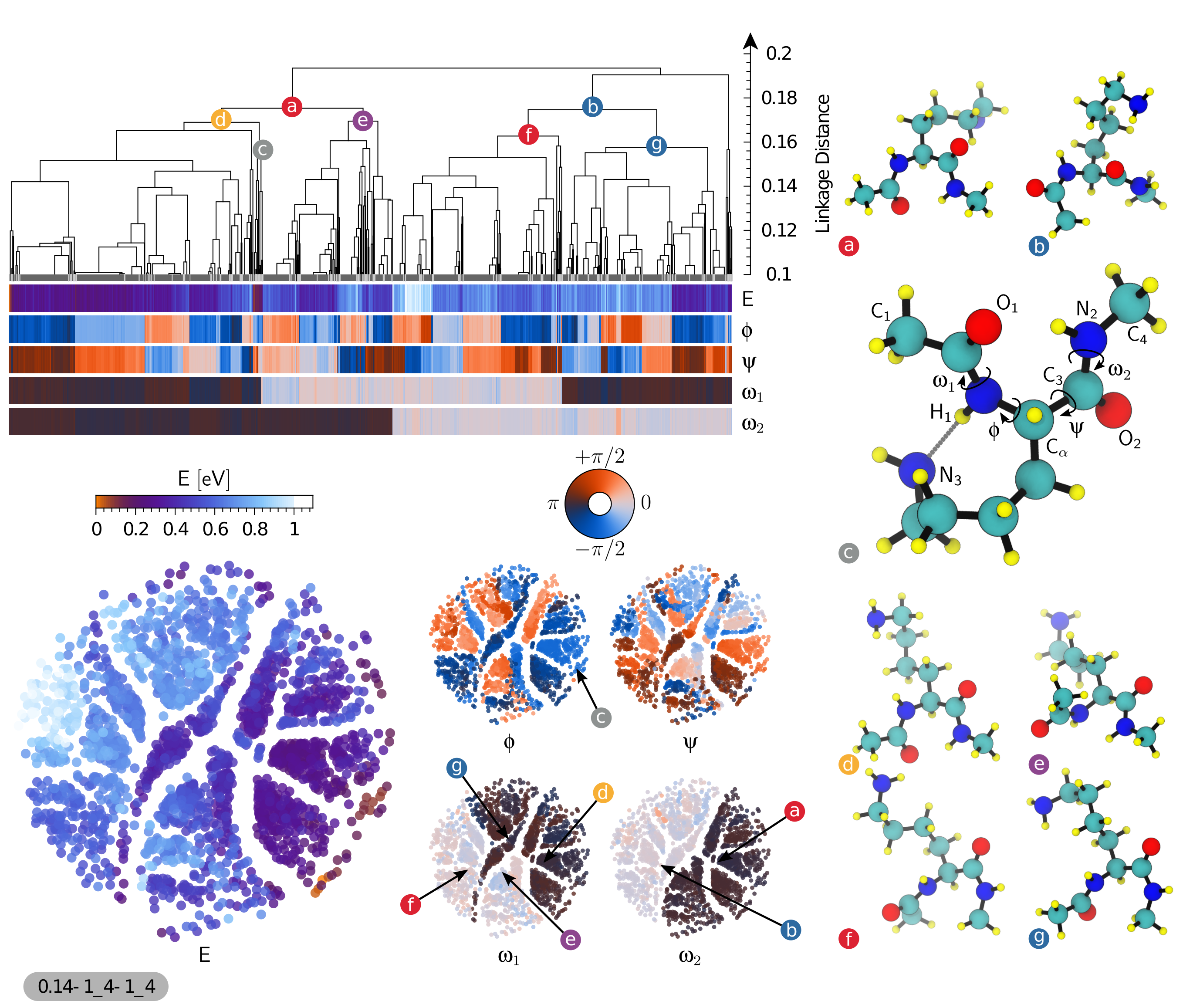} 
       
    \caption{Representation of the similarity 
    matrix corresponding to the lysine dipeptide 
    dataset using the agglomerative clustering 
    algorithm (top) and the sketchmap algorithm 
    (bottom, projection parameters shown 
    following the scheme $\sigma$-$A$\_$B$-$a$\_$b$). A few representative 
    structures (see \cref{eq:rep-struc}) of 
    interesting clusters are shown (right) and 
    their corresponding position on the 
    sketchmaps and dendrogram representation is 
    highlighted. The five sketchmaps are colored 
    according to the conformational energy and 
    the backbone dihedral angles \rachphi, 
    \rachpsi, \rachomega{1} and \rachomega{2}. 
    The dendrogram shows the clustering hierarchy 
    of the structures of the dataset. Each 
    structure is vertically aligned with its 
    properties shown using color bars below the 
    dendrogram. The dendrogram is cut at a 
    linkage distance of $0.1$ since structural 
    properties are very similar below this 
    threshold, and the clusters that are merged 
    at this level are shown as thick gray bars 
    separated by light-gray lines.
    Clusters composed of only one 
    structure are drawn as a black line reaching
    the bottom of the dendrogram. 
    The main structural motifs of this set
    of structures are governed by the peptide 
    bond dihedral angles \rachomega{1} and 
    \rachomega{2}. The two main clusters (a) and 
    (b) are showing a global correlation with the 
    angle \rachomega{2} while the angle 
    \rachomega{1} splits them into two well 
    correlated sub-clusters (d), (e) and (f), (g) 
    respectively. The cluster (c) is highlighted 
    as an example containing `outlier' structures 
    of low conformational energy. 
    }
       \label{fig:lys-dipep}
   \end{figure*}
   
    We take as our first example a subset of the database
    containing 2080 conformers of lysine dipeptide. 
    As discussed in the previous section, we start by 
    constructing the (dis)similarity matrix using the 
    SOAP-REMatch kernel. In \cref{fig:lys-dipep} the 
    dendrogram plot as well as sketchmaps have been shown 
    along with five properties, energy and four dihedral 
    angles, using the same color scales in both the 
    sketchmap and dendrogram representations. In the 
    sketchmap each circular `disk' represents a conformer. 
    Whereas in the case of the dendrogram plot, structures are 
    represented by vertical lines at the bottom of the plot.
    The strong correlation between energy and 
    conformational parameters on one side, and clustering
    and position on the map on the other, testifies how the
    the REMatch-SOAP kernel induces a meaningful
    classification of the structures in this dataset.
    
   While both clustering and sketchmap show clearly that 
   the dataset is composed of groups of structurally-related 
   conformers, the agnostic nature of the underlying metric does
   not disclose immediately the structural features that most
   transparently differentiate between different clusters. 
   Comparing the representative structures from the 
   main clusters allowed us to quickly identify 
   candidate structural motifs that could be used
   to rationalize the layout of the conformational 
   landscape. By color-coding the dendrogram
   and the sketchmaps according to these indicators
   one can readily highlight the key correlations. 
When considering existing literature on the stability of oligopeptides, the two structural parameters that are most often considered as the key coordinates to navigate the conformational landscape are the Ramachandran dihedral angles \textmyfont{\straightphi} and  \textmyfont{\textpsi}, that determine the structure of the backbone around the side chain bearing C\textalpha{} atom~\cite{ramachandran-plot} under the assumption of peptide bonds being solely in \textit{trans} conformation.
While fine-grained clusters are clearly homogeneous with respect to the \textmyfont{\straightphi} and \textmyfont{\textpsi} angles, it is clear that for the present systems they are not those that determine the clear-cut branching at the top of the dendrogram. 
An analysis of the representative structures for the two main clusters (a) and (b) shows that the two molecules differ by the isomerization of the N-terminal peptide bond.
Further splitting of these two clusters, i.e. (a) into clusters (d) and (e), and (b) into (f) and (g), depends on the isomerization of the C-terminal peptide bond. 
We can confirm this attribution of the main features of the dataset by color-coding the map and the dendrogram following the dihedral angles \rachomega{1} and \rachomega{2}. 
The four main clusters are largely homogeneous with respect to peptide bond isomerization, and are then further subdivided based on \textmyfont{\straightphi} and  \textmyfont{\textpsi}.
This observation deserves some further comment. 
Peptide bonds in naturally-occurring proteins are believed to 
almost exclusively exist in \textit{trans} conformation with the 
exception of prolyl peptide bonds where a smaller energy difference 
to \textit{trans} increases the chance for \textit{cis} 
onformers\cite{fischer_chemical_2000,dugave_cistrans_2003}.
This view is supported by the analysis of protein structures 
deposited in the protein databases where \textit{cis} 
conformations are found for about 5\% of the prolyl peptide bonds, 
but less than 0.1\% for the others\cite{weiss_peptide_1998}.
X-ray crystallographic structure represent however merely frozen snapshots
of structural dynamics. The \textit{ab initio} structure search protocol,
instead, does consider the peptide bond torsions as variable and intentionally
allowed simulations to overcome the isomerization barrier.
Consequently, the dataset contains representatives of all four 
combinations of \emph{cis} and \emph{trans} conformers. 
Since these transitions are strongly bimodal, and reflect in 
significant changes of the favorable side chain conformations, they constitute 
the most significant feature to classify the conformers. 
As expected, the most stable conformers are largely in a \emph{trans}-\emph{trans} conformation. However, the large 
parts of conformational space of that is occupied by conformers with 1 or 2 
\textit{cis} peptide bonds suggests that   \emph{cis} isomers
might play a role in the dynamics of peptides and proteins.
If the analysis had been performed solely under the assumption of the importance of the Ramachandran \textmyfont{\straightphi} and \textmyfont{\textpsi} dihedrals, it would have missed one of the main features of the structural landscape that is critical to characterize the relation between structure and energetics. 
One could then proceed further with the analysis, focusing for instance on small clusters containing low-energy structures such as that represented by the conformer (c). 
All the structure in this group are \emph{trans}-\emph{trans} isomers, that in addition have $\rachphi{}\approx -90$ degrees and $\rachpsi{}\approx 90$ degrees, allowing for the formation of a H-bond between the side chain \myatm{N}{3} and \myatm{H}{1}, and a favorable arrangement of the \myatm{N}{2} donating a H-bond to the carbonyl \myatm{O}{1} as shown in ~\cref{fig:lys-dipep}. 
Having access to the combined information on energetics, and on the grouping of structures with similar geometry makes it easier to rationalize the energy ordering of the structures, without having to separately juxtapose all of the low-lying conformers but focusing on a few representative structures. 

\subsubsection{Protonated Lysine Dipeptide}

\begin{figure*}[!ht]
    \centering
    \includegraphics[width=0.9\textwidth]{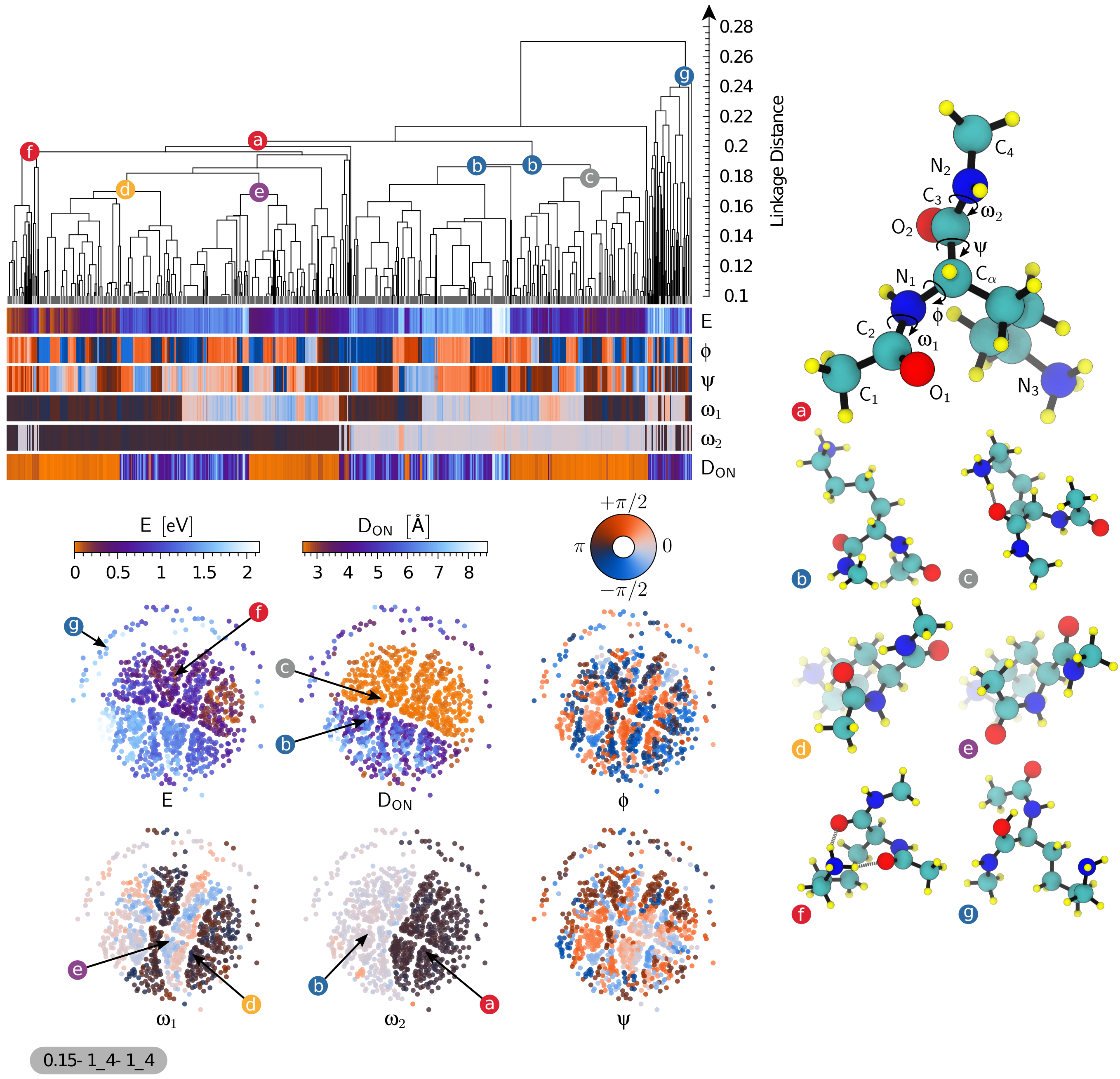} 
    \caption{Representation of the similarity matrix corresponding to the protonated lysine dipeptide dataset using the agglomerative clustering algorithm (top) and the sketchmap algorithm (bottom, projection parameters shown 
    following the scheme $\sigma$-$A$\_$B$-$a$\_$b$). A few representative structures (see \cref{eq:rep-struc}) of interesting clusters are shown (right) and their corresponding position on the sketchmaps and dendrogram representation is highlighted. 
    The six sketchmaps are colored according to the conformational energy, the minimal distance between \myatm{O}{1} or \myatm{O}{2} with \myatm{N}{3} called \myfontsp{D}{ON}, and the backbone dihedral angles \rachphi, \rachpsi, \rachomega{1} and \rachomega{2}. The dendrogram shows the clustering hierarchy of the structures of the dataset. Each structure is vertically aligned with its properties shown using color bars below the dendrogram. The dendrogram is cut at a linkage distance of $0.1$ since structural properties are very similar below this threshold, and the clusters that are merged at this level are shown as thick gray bars separated by light-gray lines. Clusters composed of only one structure are drawn as a black line reaching the bottom of the dendrogram. The main structural motifs of this set of structures are governed by the dihedral angles \rachomega{1} and \rachomega{2} and the distance \myfontsp{D}{ON}. The two main clusters (a) and (b) are showing a global correlation with the angle \rachomega{2}  while the angle \rachomega{1} splits them into well correlated sub-clusters (e.g. sub-clusters (d) and (e)). The other important sub-clustering parameter is the distance \myfontsp{D}{ON}, e.g. sub-clusters (c) and (b), which also correlates well with the separation between low and high conformational energy shown on the sketchmaps. Two sub-clusters are particular: (g) is a clear `outlier' due to a chemical change and (f) features a H-bonding pattern with the side chain NH$_3^+$ pointing to both carboxy groups that sets this cluster apart from all others.
    \label{fig:hcluster-lysH}}
\end{figure*}  
   
As the second example we considered a dataset containing 897 conformers of gas phase protonated lysine dipeptide. 
We follow the same steps as described in the previous example in order to find the most basic structural motifs for this system. 
\cref{fig:hcluster-lysH} shows the dendrogram, the sketchmap and a few color coded properties for this system to show their correlation with the classification. 
The most prominent feature for this molecule, which is evident in both the dendrogram and the sketch maps, is the presence of a group of outliers, that are clearly separated from the bulk of the conformers. 
Inspection of the cluster centroid (g) clarifies the structural basis of this separation.
Whereas in most of the structures the excess charge lies on the lysine side chain as a NH$_3^+$ group, conformers in this cluster experienced a proton transfer event, with the excess proton attached to one of the carbonyl oxygen \myatm{O}{1}, stabilized by H-bonding to \myatm{N}{2}.
This is a result of the database generation where \textit{ab initio} replica-exchange molecular dynamics including high T trajectories where used for structure sampling during which protons can eventually transfer.

Moving on to the main cluster of structures, we can see that similar to our previous example of the neutral dipeptide and again due to the unbiased sampling protocol and the high energy range
the peptide bond angles are again more important than Ramachandran's dihedrals. 
Conformers (a) and (b) are the  representative structure for groups having \textit{cis} and \textit{trans} \rachomega{2} peptide bonds respectively. 
Group (a) is further split based on the \textit{cis/trans} state of \rachomega{1} into the clusters represented by structures (d) and (e).

The presence of a charged side chain leads to stronger H-bonds and to less clear-cut grouping based on peptide-bond isomerism than for the neutral dipeptide.
This is apparent for instance from the presence of subclusters such as that represented by conformer (f), in which the bent side chain leads to the formation of two H-bonds between NH$_3^+$ group and the carbonyl oxygens. 
H-bonds also dominate the partitioning of cluster (b), that is split into two groups -- one of which is still best represented by the same conformer, and one that is epitomised by (c). 
Once again, inspection of these structural representatives reveals the organising principle behind the classification:
(c)-like structures have an extended side chain, and are dominated by interactions among the peptide bond moieties, whereas (b)-like structures have a well-formed H-bond between the side chain and one of the two backbone O atoms. 
This structural pattern can be emphasized by color-coding conformers based on the parameter \myfontsp{D}{ON}$=\min \left[ d(\myatm{O}{1},\myatm{N}{3}); d(\myatm{O}{2},\myatm{N}{3})\right]$:
A small O-N distance indicates bending of the side chain and the formation of a H-bond between O and N.
As it is clear from the sketchmap representation, there is a very strong correlation between the bending of the charged side chain and the energy of a conformer, with all of the structures within 0.5~eV of the ground state feature this sidechain to backbone H-bonds.

A scenario with such a clear-cut partitioning is less likely to happen with oligopeptides in a polar solvent like water, where intramolecular H-bonds that introduce strain  compete with 
H-bonds with the surrounding water molecules, without the need of bending the structure.
The analysis techniques we introduce in this work would be ideally suited to rationalize the changes in the (free) energetics of biological molecules when moving from the gas phase to (micro)solvated environments or to organic/inorganic interfaces.

\subsubsection{Uncapped Lysine}
   
\begin{figure*}[!ht]
    \centering
    \includegraphics[width=\textwidth]{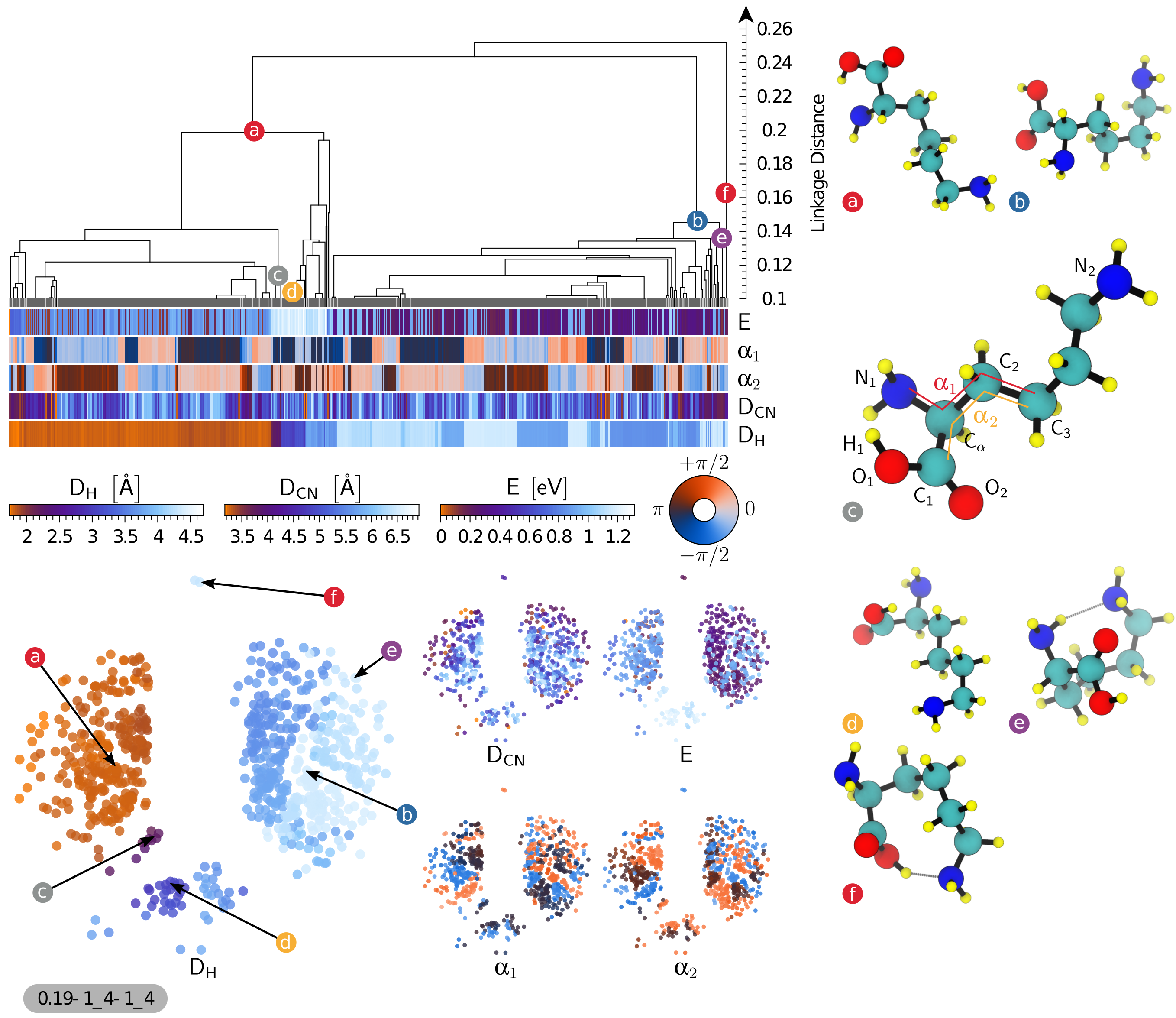}
    \caption{Representation of the similarity matrix corresponding to the lysine uncapped dataset using the agglomerative clustering algorithm (top) and the sketchmap algorithm (bottom, projection parameters shown 
    following the scheme $\sigma$-$A$\_$B$-$a$\_$b$). A few representative structures (see \cref{eq:rep-struc}) of interesting clusters are shown (right) and their corresponding position on the sketchmaps and dendrogram representation is highlighted. The five sketchmaps are colored according to the conformational energy, the distance between \myatm{N}{1} and the hydrogen in the carboxilic group \myatm{H}{1} (labelled \myfontsp{D}{H}), the distance between \myatm{N}{2} and \myatm{C}{\alpha} (labelled \myfontsp{D}{CN}), and the dihedral angles \rachalpha{1} and \rachalpha{2} which are respectively computed with the following atoms  (\myatm{N}{1},\myatm{C}{\alpha},\myatm{C}{2},\myatm{C}{3}) and (\myatm{C}{1},\myatm{C}{\alpha},\myatm{C}{2},\myatm{C}{3}). The dendrogram shows the clustering hierarchy of the structures of the dataset. Each structure is vertically aligned with its properties shown using color bars below the dendrogram. The dendrogram is cut at a linkage distance of $0.1$ since structural properties are very similar below this threshold, and the clusters that are merged at this level are shown as thick gray bars separated by light-gray lines. Clusters composed of only one structure are drawn as a black line reaching the bottom of the dendrogram. The main structural motifs of the database are governed by the distance \myfontsp{D}{H}. The two main clusters (a) and (b) are agglomerated according to the orientation of \myatm{H}{1} and the oxygen atom it is bonded to with respect to \myatm{N}{1} which is well described by the distance \myfontsp{D}{H}. The sub-cluster (e) is composed of `outlier' structures showing an H-bond between \myatm{N}{2} and an hydrogen of \myatm{N}{1} resulting in a folded side chain structural motif. 
    Finally, the outlier cluster (f) contains a H-bond between 
    the carboxy H and the side-chain \myatm{NH}{2}, that
    can be seen as a precursor to the zwitterionic form.}
    \label{fig:hcluster-lysuncapped}
\end{figure*}  

Our third example is a dataset containing 733 conformers of the un-capped lysine molecule in the gas phase. 
We follow the same steps as described in the previous examples to construct the dendrogram shown in \cref{fig:hcluster-lysuncapped}. 
The map has a simple structure, with few well-separated groups. 
Being a smaller molecule with fewer degrees of freedom, the Ramachandran angles are not defined.
Still, the dihedral angles in the vicinity of the \myatm{C}{\alpha} atom display local structural correlation but once again they are not the main organizing factor that can rationalize the clustering. 
By juxtaposing representative conformers from the main clusters we could identify a better order parameter, that correlates strongly with H-bond patterns within the molecule. 
Namely, the distance (\myfontsp{D}{H}) between the H atom in the \myatm{OH}{} group of the carboxy function and the \myatm{N}{} atom in the backbone (\myatm{N}{1}) discriminates well between structures based on H-bonding patterns\cite{ropo_scirep_2016} of \textit{type~I} between \myatm{N}{1}\myatm{H}{}$\rightarrow$\myatm{O}{2} (e.g. conformer (b)) and of \textit{type~II} with a H-bond \myatm{O}{1}\myatm{H}{}$\rightarrow$\myatm{N}{1} (e.g. conformer (a)). 
It can  be seen from both the dendrogram and the sketchmaps that one could identify several further  subgroups  based on particular values of \myfontsp{D}{H}, representing specific orientations. 
Conformers (c) and (d) represent  small groups of conformers having specific relative orientation between the OH and \myatm{NH}{2} groups. 
Conformer (e) is representative of a small outlier group with a well-defined bend of the side chain, leading to the formation of a further H-bond between the \myatm{N}{1} atom in the amino acid moiety and \myatm{N}{2}, in the side chain.
The lysine side chain is very flexible, and the distance between \myatm{N}{} and \myatm{C}{\alpha} only plays a role in defining the fine-grained structure of the dataset, but is minimally correlated with the most prominent features.

 \begin{figure*}[!htbp]
       \centering
        \includegraphics[width=0.8\textwidth]{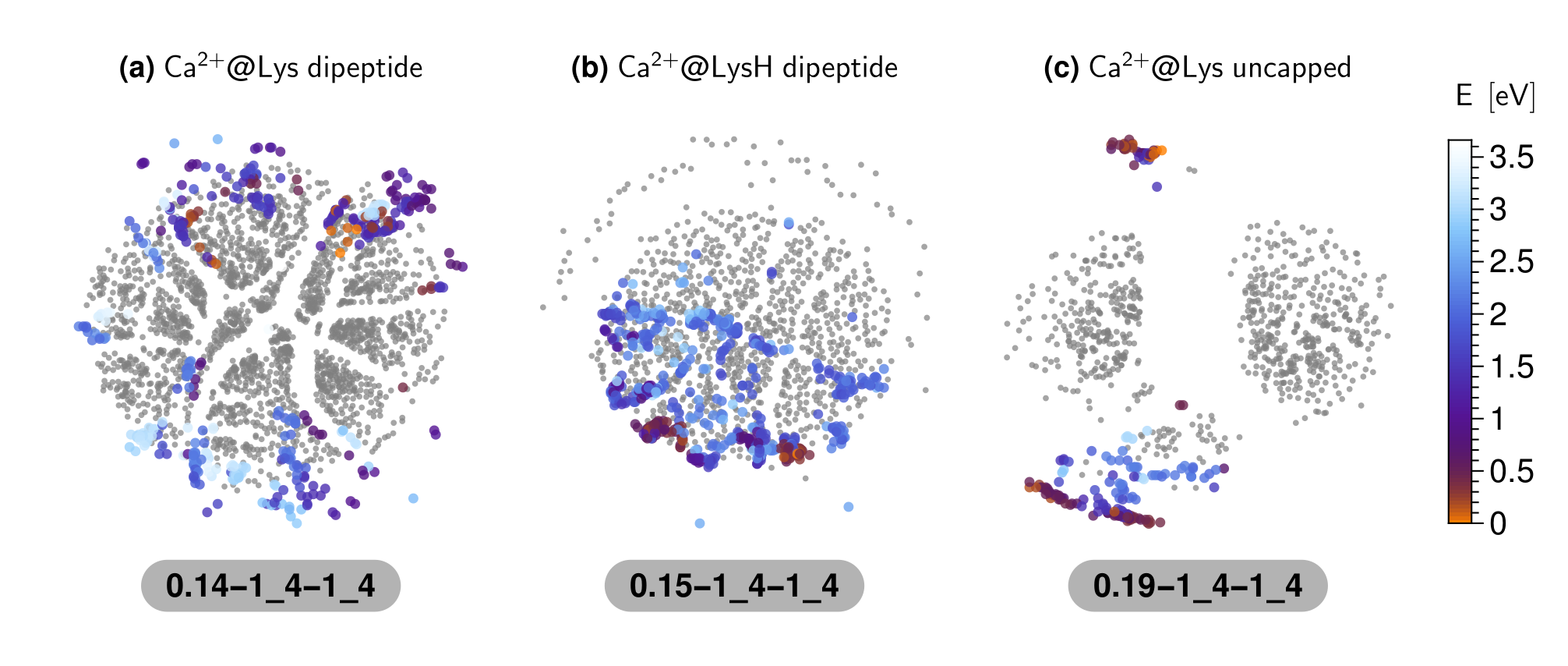}
       \caption{The out-of-sample embedding of conformers with \ce{Ca^2+} ion on the sketchmap of their pure counterpart, for the three systems we discussed in above: lysine dipeptide (a), protonated lysine dipeptide (b) and molecular lysine (c) systems. The projected conformers are colored with their energy where as the sketchmap on which they are projected are kept all in grey color. The location of the projected conformers allows us to understand how the conformational space of the pure conformers are affected due to presence of the \ce{Ca^2+} ion. 
       }
       \label{fig:smap-Ca}
   \end{figure*} 
   
While it appears that even in this case we could identify the basic structural motifs that characterize the conformational landscape of this system, the correlation with energy is very poor.
There are several instances in both the dendrogram and the sketchmap where two conformers that are detected as structurally very similar while they display very different stability. 
Understanding whether this inconsistency signals a problem with our analysis brings us to the topic of outlier detection and consistency checks, that we will discuss in details in Section~\ref{sec:outliers}. 

\subsection{Understanding the Impact of Perturbations
on conformational Space}
Having elucidated the essential structural motifs that underlie the organization of a set of molecular conformers, one could also wonder how changes in the thermodynamic conditions, or another external perturbation such as solvation, the addition or subtraction of an electron~\cite{Dejcp11} or that of an atom~\cite{ideh_mg,ghazi_Na,pascalB80}, modify the conformations of the molecule and their stability.
The database\cite{db-paper,aminodb} that we are using here as an example contains, in addition to bare oligopeptides, sets of locally-stable conformers in the presence of cations of six different species, namely \ce{Ca^2+}, \ce{Ba^2+}, \ce{Sr^2+}, \ce{Cd^2+}, \ce{Pb^2+} and \ce{Hg^2+}. 
We take the example of \ce{Ca^2+} to describe how one can characterize its impact on the conformational space of the three molecular systems that we have discussed in our previous examples.
We start by calculating the dissimilarity of all the conformers containing cations with their pure counterpart.
In order to make the comparison on the same footings, we did not include the location of the cation in defining the SOAP kernels, so that the presence of \ce{Ca^2+} only enters by distorting molecular geometries and/or altering their relative stability.
Using this information, we then projected the cation-containing dataset on the top of the sketchmap of structures for the bare molecule. 
This is done using sketchmap out-of-sample embedding, and we refer our reader to see the relevant literature~\cite{ceri+11pnas,trib+12pnas,ceri+13jctc} for more details about the method. 
In \cref{fig:smap-Ca} we show the resulting projection, colored according to the stability of the conformers, on top of the sketchmap of the pure molecule shown in grey color as a reference. 
A close proximity of projected conformers with a pure conformer signifies their structural similarity.
Segregation of the projected conformers with the cation in some area of the reference sketchmap, represents the structural bias introduced by the strong electrostatic interaction with \ce{Ca^2+}. 
 
In the case of neutral lysine dipeptide (\mysubfigref{fig:smap-Ca}{a}), the presence of the Ca$^{2+}$ ion induces overall relatively small distortions of the stable conformers, that get pushed towards the outer region of the map but are still clearly related to the  locally stable structures for the bare molecule.
Energies are dramatically changed, with the most stable cluster in the original map begin completely absent in the presence of the cation. 
These observations highlight the importance of sampling high-energy conformers during high-throughput structure searches, since the relative stability can be modulated strongly by external perturbations.  In particular, \textit{cis} conformers become energetically more competitive  
and are topologically closer to the global minima.
In the case of protonated lysine dipeptide  (\mysubfigref{fig:smap-Ca}{b}), the same analysis
shows an even clearer pattern.
All the conformers with \ce{Ca^2+} ions are projected in the lower part of the  sketchmap, that correspond to conformers with an extended side chain (see \cref{fig:hcluster-lysH}). 
The \ce{Ca^2+} ion preferably binds to the peptide \myatm{O}{} atoms, and the electrostatic repulsion with the protonated lysine residue strongly favors extended conformers, contrary to what observed in the case of the bare molecule.
Finally, one sees that for molecular lysine (\mysubfigref{fig:smap-Ca}{c}) the addition of Ca$^{2+}$ leads to conformers with very different structural motifs from those seen with the bare molecule, which is apparent in the sketchmap projection being concentrated far away from the unperturbed conformers. 
In fact, inspection of the structures shows that in most cases Ca$^{2+}$ triggers the transition to the zwitterionic form, with the cation coupled to the carboxylate group, and the side chain protonated NH$^3_+$ extending as far as possible away from it. 
In analogy with what observed for Lennard-Jones clusters~\cite{ceri+13jctc} and solvated polypeptide segments~\cite{arde+15jctc}, sketchmaps proved to be a powerful tool to analyze the response of the system to external perturbations and changes in the boundary conditions, and -- in this specific example -- do draw connections between different subsets of a high-throughput molecular database. 

\subsection{Identifying Outliers and Checking for Consistency}
\label{sec:outliers}
The tools we introduced in this work are useful to 
address other important issues in data-driven science, 
such as outlier detection and consistency checks. 
We have already discussed the importance of detecting 
groups of ``outlier'' structures that are very different 
from the bulk of the dataset. 
These unusual items often signal the occurrence of 
unexpected effects that go beyond the original goal 
of the database construction effort. In the case
of protonated lysine dipeptide, looking for outliers
allowed us to reveal the presence of conformers with 
different chemical connectivity, or of strong H-bonds
between the backbone and the charged side chain. 
Similar observations can also be made in the case of
the bare lysine molecule (\cref{fig:hcluster-lysuncapped}).
Also in this case, one can observe a branch at the topmost 
level of the dendrogram, containing only two conformers. 
These are the only two cases where a H-bond is formed 
between the \myatm{N}{} of the side chain and the 
\myatm{H}{} atom of the \myatm{OH}{} group in the backbone. 
In the sketchmap, these two conformers are projected on 
the top, clearly isolated from rest of the groups, and bear
the most resemblance to the zwitterionic conformers that are 
stabilized in the presence of a divalent cation.
Obviously, the definition of a group of ``outliers''
can be more nuanced, and refer to small groups of 
structures appearing at deeper levels in the hierarchy.
Overall, the possibility of clustering together the
structures in a large dataset and inspect a few
representative conformers, rather than hundreds or thousands,
greatly facilitates the task of identifying trends and
spotting interesting or unexpected structures.

\begin{figure}
    \centering
    \includegraphics[width=0.5\textwidth]{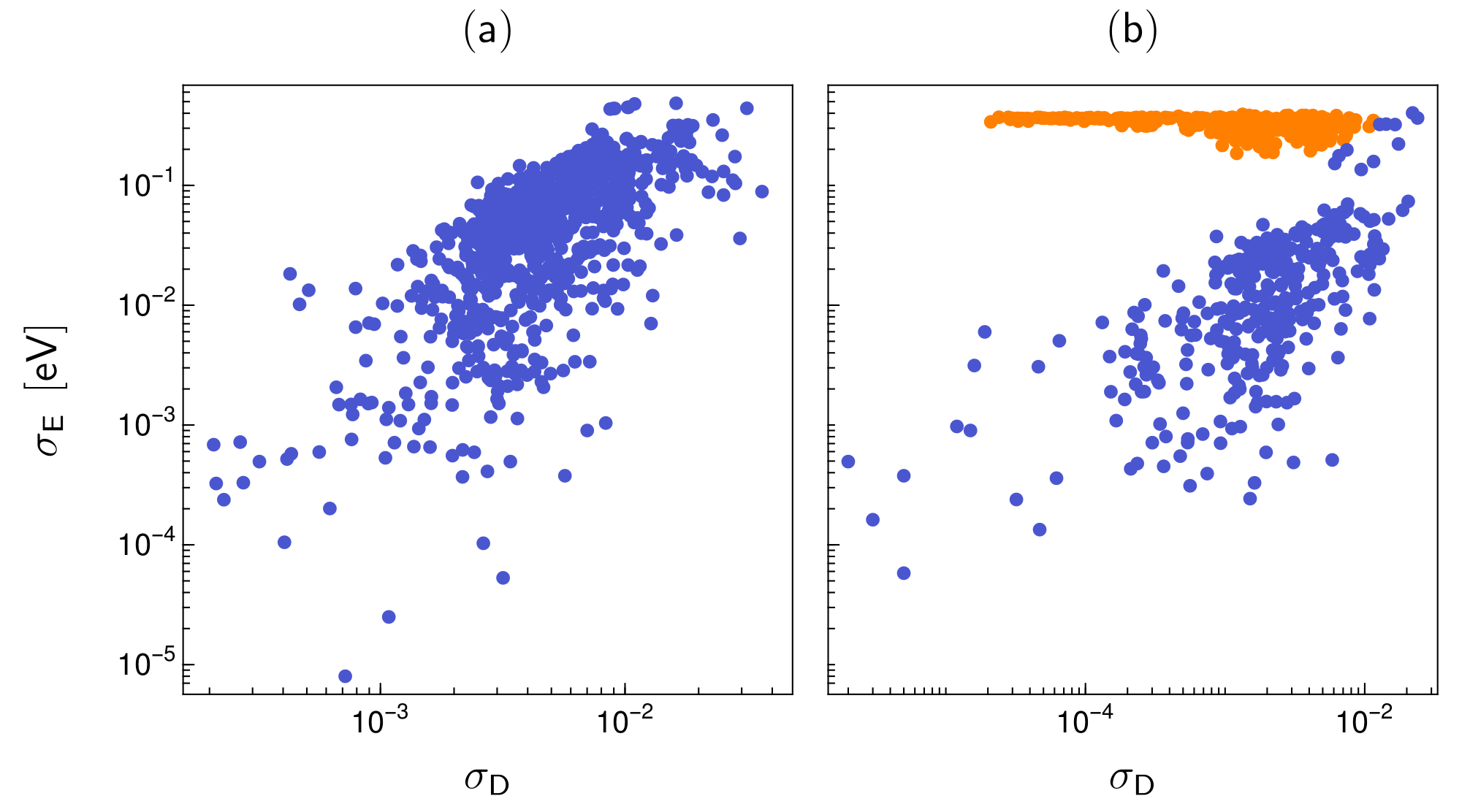}
    \caption{This figure compares the homogeneity of clusters from the protonated lysine dipeptide (see a) and the bare lysine uncapped (see b) with respect to properties of their elements. The homogeneity of a cluster is probed using the standard deviation with respect to the distance between each cluster elements, \textsigma$_D$, and the conformational energy, \textsigma$_E$. The outliers of uncapped lysine (b) were manually highlighted in orange.}
    \label{fig:consistency}
\end{figure} 

Outliers can signal interesting or important 
trends, but can also be a red flag for the presence 
of errors. The importance of database integrity 
has long been recognized by computer 
scientists~\cite{dataint1,dataint2,dataint3,dataint4}, 
and several tools are available to monitor and 
correct inconsistencies from the technical 
point of view, in terms of reliability of storing 
and retrieving data\cite{outlier1,outlier2,outlier3,outlier4,outlier5,outlier6}. 
The issue is also crucial when it comes to the maintenance
of automatically-generated databases, and to repositories
that store data of heterogeneous provenance\cite{aiida,clean_en,esp,oqmd,pauling,matgenome1,matgenome2}. 
In these cases, problems have generally little to do with the integrity of the storage, but rather with the consistency of the simulation details of different sets of calculations.
Inconsistencies should here manifest themselves in the presence of structures that are geometrically very similar, but are associated to very different values of particular properties. 

For example the lysine molecule dataset shows signs of this kind
of issues, with energies that vary wildly within clusters that are 
very homogeneous in structure.
This problem can be seen from the maps, i.e. when comparing the 
energy-colored sketchmap in \cref{fig:hcluster-lysuncapped} 
to the respective maps for the other systems.
However, a more robust and easy-to-automate approach to identify 
structure/property inconsistencies starts from the hierarchical 
clusters, and compares the structural variability within each 
cluster $\sigma_D$ (\cref{eq:sigma-x}) with the variance 
of a given property, in this case energy, $\sigma_E$. 
Looking, for example, at a glassy energy landscape\cite{De2011}, one 
can observe configurations that are very different from a 
structural point of view, but have similar energy, giving 
rise to clusters with large $\sigma_D$ and small $\sigma_E$. 
The data points in \cref{fig:consistency} each represent 
individual clusters of lysine dipeptide and uncapped lysine, 
respectively, and illustrate their variation in energy and structure. 
In the case of lysine dipeptide (\cref{fig:consistency}a) one sees a clear
correlation between the structural and energetical variation of the clusters.
The two quantities $\sigma_D$ and $\sigma_E$ are not necessarily strongly 
correlated, but in general clusters that contain very similar structures 
also have a low spread in energy.
For uncapped lysine (\cref{fig:consistency}b), however,
one can identify a group of points 
(which we manually highlighted in orange for clarity)
that has a distinctively different behavior, with
$\sigma_E$ converging to a constant value other
than zero as $\sigma_D$ decreases. 
This kind of feature indicates that the metric that 
is used to classify structures cannot detect one 
specific effect that has a dramatic impact on 
energetics, signaling either a failure of the 
metric or, as in this case, an inconsistency in 
the generated data.
Further investigation of the lysine 
molecule dataset revealed that a subset of 
structures that had been generated 
at a lower level of theory in the initial
stages of the structure-search procedure
made their way by mistake into the final dataset. 
Using this measure of cluster homogeneity 
on all systems of the amino acid database 
(see Supporting Information) 
revealed similar problems also for other 
molecules, for example Cys, Glu, and Arg.
Thanks to this analysis we will be able to 
identify and rectify mistakes in all the affected 
datasets and subsequently update the on-line 
repository\cite{aminodb}.

\section{Conclusion}

The increasing use of high-throughput 
computational screening of materials and
molecules, and the compilations of more-or-less
curated databases of the resulting structures and properties, 
is making more and more urgent to 
adapt ``big data'' techniques to the 
problems that are specific to this field. 
In this work we have demonstrated how 
a toolbox of algorithms ranging from 
hierarchical clustering to non-linear 
dimensionality reduction can be used to
navigate molecular databases, taking as a 
paradigmatic example some subsets of a 
database of oligopeptide structures in the 
gas phase. The software that was used
to compute similarity data between molecules, 
as well as to generate 
dendrograms and sketch-maps, are open-source and
available for download\cite{cosmogit,libatoms}.

We find that the use of REMatch-SOAP, 
a general and unbiased
metric to compare different structures 
based on the combination of pair-wise 
similarity between molecular environments,
makes these techniques particularly 
insightful.  Rather than simply reflecting 
preconceived notions of what would be the
key structural parameters to differentiate
molecular conformers, this metric reveals 
for instance the importance of peptide bond
isomerization in describing the high-energy
portion of conformational space of 
oligopeptides, the possibility of changes in
chemical connectivity in the course of the
\textit{ab initio} structural search, and the interplay
between hydrogen-bonding, backbone dihedrals
an electrostatic interactions. 
Sketchmaps and hierarchical clustering 
proved to be complementary tools, with 
representative structures from the main 
clusters providing an easy way to compare
visually different groups of conformers, and 
the low-dimensional map providing a quick,
intuitive tool to verify hypotheses and 
visualize structure-property correlations.
   
Assumption-free first-principles molecular-structure 
search for data generation 
in combination with dimensionality reduction and 
clustering for data analysis provide a powerful tool box 
for studying structure formation trends.
We could highlight the presence of large portions
of configurational space that consist of \emph{cis}
isomers of the peptide bond. Albeit energetically 
unfavorable, these conformers may play an important
role in the dynamics of polypeptides. 
By comparing isolated molecules with their
complexes with \ce{Ca^2+}, we can also reveal
how a strong electrostatic perturbation modifies 
the energetic landscape of a 
small molecule -- be it by shifting the 
stability of different conformers, or 
triggering the formation of new structures
that are absent in the absence of a cation. 
Furthermore, we also demonstrate the importance of automated analysis techniques
in assessing the integrity and internal consistence of a database, 
by successfully identifying a subset of structures associated with 
ill-converged energetics. 

All of the techniques we discussed 
should be readily extendable to
heterogeneous databases of molecules and 
solids, where we expect that the possibility
of defining an alchemical kernel within the 
REMatch-SOAP metric will make it possible to
tune the relative weight of composition and
structure in determining the notion
of similarity. 
By simplifying the analysis and the 
interpretation of computational datasets
containing thousands or millions of 
hypothetical compounds, these methods will
be crucial to unleash the full potential
of computational materials design. 

\section{Declarations}

\subsection{Availability of data and materials}
The oligopeptide database used as an example in this paper is already available online ~\cite{aminodb}.
All the data required to generate the figures in this paper and same analysis data for other oligopeptides in the 
database are provided in the supplementary information. Similarity matrix data is not included in SI due to size 
limitations, but  is freely available from from our group repository ~\cite{cosmogit}.

\subsection{Competing interests}
   We confirm that none of the authors have any competing interests in the manuscript.
   
\subsection{Funding}
 S.D. and M.C. would like to acknowledge support from 
the NCCR MARVEL. M.C., T.I. and C.B. would like to 
acknowledge funding from the MPG-EPFL center 
for molecular nanoscience.

\subsection{Authors' contributions}

M.C. and  S.D. designed the calculations and 
developed the methods.
F.M. and S.D. performed calculations and analysis,
and prepared the materials for the manuscript. 
C.B. and T.I.  provided insights on the implications
of the analysis of the oligopeptide database. 
All the authors contributed to the writing of the
manuscript.

\subsection{Acknowledgements}
S.D. would like to thank Czuee Morey (University of Geneva) for insightful discussion.   
C.B. thanks Matti Ropo (Tampere University of Technology) 
Volker Blum (Duke University) and Matthias Scheffler 
(Fritz Haber Institute) for support and discussion.


\begin{thebibliography}{10}

\bibitem{aiida}
Pizzi G, Cepellotti A, Sabatini R, Marzari N, Kozinsky B.
\newblock {AiiDA: automated interactive infrastructure and database for
  computational science}.
\newblock Computational Materials Science. 2016;111(1):218--230.

\bibitem{clean_en}
Hachmann J, Olivares-Amaya R, Atahan-Evrenk S, Amador-Bedolla C,
  S{\'{a}}nchez-Carrera RS, Gold-Parker A, et~al.
\newblock {The harvard clean energy project: Large-scale computational
  screening and design of organic photovoltaics on the world community grid}.
\newblock Journal of Physical Chemistry Letters. 2011;2(17):2241--2251.

\bibitem{esp}
Ortiz C, Eriksson O, Klintenberg M.
\newblock {Data mining and accelerated electronic structure theory as a tool in
  the search for new functional materials}.
\newblock Computational Materials Science. 2009;44(4):1042--1049.

\bibitem{oqmd}
Saal JE, Kirklin S, Aykol M, Meredig B, Wolverton C.
\newblock {Materials Design and Discovery with High-Throughput Density
  Functional Theory: The Open Quantum Materials Database (OQMD)}.
\newblock JOM. 2013;65(11):1501--1509.

\bibitem{pauling}
Villars P, Berndt M, Brandenburg K, Cenzual K, Daams J, Hulliger F, et~al.
\newblock {The Pauling File, Binaries Edition}.
\newblock Journal of Alloys and Compounds. 2004;367(1-2):293--297.

\bibitem{matgenome1}
Jain A, Ong SP, Hautier G, Chen W, Richards WD, Dacek S, et~al.
\newblock {Commentary: The materials project: A materials genome approach to
  accelerating materials innovation}.
\newblock APL Materials. 2013;1(1):011002.

\bibitem{matgenome2}
White A.
\newblock {The Materials Genome Initiative: One year on}.
\newblock MRS Bulletin. 2012;37(08):715--716.

\bibitem{PhysRevLett.108.058301}
Rupp M, Tkatchenko A, M{\"{u}}ller KR, von Lilienfeld OA.
\newblock {Fast and Accurate Modeling of Molecular Atomization Energies with
  Machine Learning}.
\newblock Physical Review Letters. 2012;108(5):058301.

\bibitem{PhysRevLett.114.105503}
Ghiringhelli LM, Vybiral J, Levchenko SV, Draxl C, Scheffler M.
\newblock {Big data of materials science: Critical role of the descriptor}.
\newblock Physical Review Letters. 2015;114(10):105503.

\bibitem{PhysRevB.92.014106}
Huan TD, Mannodi-Kanakkithodi A, Ramprasad R.
\newblock {Accelerated materials property predictions and design using
  motif-based fingerprints}.
\newblock Physical Review B - Condensed Matter and Materials Physics.
  2015;92(1):14106.

\bibitem{PhysRevB.92.094306}
Botu V, Ramprasad R.
\newblock {Learning scheme to predict atomic forces and accelerate materials
  simulations}.
\newblock Physical Review B - Condensed Matter and Materials Physics.
  2015;92(9):94306.

\bibitem{kusne15screport}
Kusne A, Gao T, Mehta A, Ke L, {Cuong Nguyen} M, Ho KM, et~al.
\newblock {On-the-fly machine-learning for high-throughput experiments: search
  for rare-earth-free permanent magnets}.
\newblock Scientific Reports. 2014;4:6367.

\bibitem{ramkrsinan_2014sd}
Ramakrishnan R, Dral PO, Rupp M, von Lilienfeld OA.
\newblock {Quantum chemistry structures and properties of 134 kilo molecules.}
\newblock Scientific data. 2014;1:140022.

\bibitem{PhysRevB.90.155136}
Arsenault LF, Lopez-Bezanilla A, {Von Lilienfeld} OA, Millis AJ.
\newblock {Machine learning for many-body physics: The case of the Anderson
  impurity model}.
\newblock Physical Review B - Condensed Matter and Materials Physics.
  2014;90(15):155136.

\bibitem{db-paper}
Ropo M, Schneider M, Baldauf C, Blum V.
\newblock {First-principles data set of 45,892 isolated and cation-coordinated
  conformers of 20 proteinogenic amino acids}.
\newblock Scientific Data. 2016;3:160009.

\bibitem{rodr-laio14science}
Rodriguez A, Laio A.
\newblock {Clustering by fast search and find of density peaks}.
\newblock Science. 2014;344(6191):1492--1496.

\bibitem{clustering-rui}
Xu R, Wunsch D.
\newblock {Survey of clustering algorithms}.
\newblock IEEE Transactions on Neural Networks. 2005;16(3):645--678.

\bibitem{gang_clustering}
Yu G, Chen J, Zhu L.
\newblock {Data mining techniques for materials informatics: Datasets preparing
  and applications}.
\newblock In: 2009 2nd International Symposium on Knowledge Acquisition and
  Modeling, KAM 2009. vol.~2; 2009. p. 189--192.

\bibitem{cartography}
Isayev O, Fourches D, Muratov EN, Oses C, Rasch K, Tropsha A, et~al.
\newblock {Materials cartography: Representing and mining materials space using
  structural and electronic fingerprints}.
\newblock Chemistry of Materials. 2015;27(3):735--743.

\bibitem{prasanna-sd}
Balachandran PV, Theiler J, Rondinelli JM, Lookman T.
\newblock {Materials Prediction via Classification Learning}.
\newblock Scientific Reports. 2015;5:13285.

\bibitem{ferg+10pnas}
Ferguson AL, Panagiotopoulos AZ, Debenedetti PG, Kevrekidis IG.
\newblock {Systematic determination of order parameters for chain dynamics
  using diffusion maps.}
\newblock Proceedings of the National Academy of Sciences of the United States
  of America. 2010;107(31):13597--602.

\bibitem{ceri+11pnas}
Ceriotti M, Tribello GA, Parrinello M.
\newblock {From the Cover: Simplifying the representation of complex
  free-energy landscapes using sketch-map.}
\newblock Proceedings of the National Academy of Sciences.
  2011;108(32):13023--13028.

\bibitem{trib+12pnas}
Tribello Ga, Ceriotti M, Parrinello M.
\newblock {Using Sketch-Map Coordinates to Analyze and Bias Molecular Dynamics
  Simulations}.
\newblock Proceedings of the National Academy of Sciences.
  2012;109(14):5196--5201.

\bibitem{ceri+13jctc}
Ceriotti M, Tribello GA, Parrinello M.
\newblock {Demonstrating the transferability and the descriptive power of
  sketch-map}.
\newblock Journal of Chemical Theory and Computation. 2013;9(3):1521--1532.

\bibitem{rohr+arpc13}
Rohrdanz MA, Zheng W, Clementi C.
\newblock {Discovering Mountain Passes via Torchlight: Methods for the
  Definition of Reaction Coordinates and Pathways in Complex Macromolecular
  Reactions}.
\newblock Annual Review of Physical Chemistry. 2013;64(1):295--316.

\bibitem{de+16pccp}
De S, Bart{\'{o}}k AP, Cs{\'{a}}nyi G, Ceriotti M.
\newblock {Comparing molecules and solids across structural and alchemical
  space}.
\newblock Phys Chem Chem Phys. 2016;18(20):13754.

\bibitem{aminodb}
Ropo M, Baldauf C, Blum V. {Berlin ab initio amino acid DB}; 2016.

\bibitem{sprint}
Pietrucci F, Andreoni W.
\newblock {Graph theory meets ab initio molecular dynamics: Atomic structures
  and transformations at the nanoscale}.
\newblock Physical Review Letters. 2011;107(8):85504.

\bibitem{PhysRevB.90.104108}
Szlachta WJ, Bart{\'{o}}k AP, Cs{\'{a}}nyi G.
\newblock {Accuracy and transferability of Gaussian approximation potential
  models for tungsten}.
\newblock Physical Review B - Condensed Matter and Materials Physics.
  2014;90(10):104108.

\bibitem{PhysRevB.89.235411}
Lopez-Bezanilla A, {Von Lilienfeld} OA.
\newblock {Modeling electronic quantum transport with machine learning}.
\newblock Physical Review B - Condensed Matter and Materials Physics.
  2014;89(23):235411.

\bibitem{pilania13screport}
Pilania G, Wang C, Jiang X, Rajasekaran S, Ramprasad R.
\newblock {Accelerating materials property predictions using machine learning}.
\newblock Scientific Reports. 2013;3:2810.

\bibitem{PhysRevB.88.054104}
Bart{\'{o}}k AP, Gillan MJ, Manby FR, Cs{\'{a}}nyi G.
\newblock {Machine-learning approach for one- and two-body corrections to
  density functional theory: Applications to molecular and condensed water}.
\newblock Physical Review B - Condensed Matter and Materials Physics.
  2013;88(5):054104.

\bibitem{rupp+07jcim}
Rupp M, Proschak E, Schneider G.
\newblock {Kernel approach to molecular similarity based on iterative graph
  similarity}.
\newblock Journal of Chemical Information and Modeling. 2007;47(6):2280--2286.

\bibitem{hirn+15arxiv}
Hirn M, Poilvert N, Mallat S.
\newblock {Quantum Energy Regression using Scattering Transforms}.
\newblock arXiv preprint arXiv:150202077. 2015;.

\bibitem{qm7b}
Montavon G, Rupp M, Gobre V, Vazquez-Mayagoitia A, Hansen K, Tkatchenko A,
  et~al.
\newblock {Machine learning of molecular electronic properties in chemical
  compound space}.
\newblock New Journal of Physics. 2013;15(9):95003.

\bibitem{PhysRevLett.108.253002}
Snyder JC, Rupp M, Hansen K, M{\"{u}}ller KR, Burke K.
\newblock {Finding density functionals with machine learning}.
\newblock Physical Review Letters. 2012;108(25):253002.

\bibitem{PhysRevB.92.045131}
Ghasemi SA, Hofstetter A, Saha S, Goedecker S.
\newblock {Interatomic potentials for ionic systems with density functional
  accuracy based on charge densities obtained by a neural network}.
\newblock Physical Review B. 2015;92(4):045131.

\bibitem{anatoleIJQC}
{Von Lilienfeld} OA.
\newblock {First principles view on chemical compound space: Gaining rigorous
  atomistic control of molecular properties}.
\newblock International Journal of Quantum Chemistry. 2013;113(12):1676--1689.

\bibitem{bag.of.bonds}
Hansen K, Biegler F, Ramakrishnan R, Pronobis W, {Von Lilienfeld} OA,
  M{\"{u}}ller KR, et~al.
\newblock {Machine learning predictions of molecular properties: Accurate
  many-body potentials and nonlocality in chemical space}.
\newblock Journal of Physical Chemistry Letters. 2015;6(12):2326--2331.

\bibitem{newstefan}
Zhu L, Amsler M, Fuhrer T, Schaefer B, Faraji S, Rostami S, et~al.
\newblock {A fingerprint based metric for measuring similarities of crystalline
  structures}.
\newblock The Journal of Chemical Physics. 2016;144(3):034203.

\bibitem{cutu13nips}
Cuturi M.
\newblock {Sinkhorn Distances: Lightspeed Computation of Optimal Transport}.
\newblock In: Burges CJC, Bottou L, Welling M, Ghahramani Z, Weinberger KQ,
  editors. Advances in Neural Information Processing Systems 26. Curran
  Associates, Inc.; 2013. p. 2292--2300.

\bibitem{pca-WOLD198737}
Wold S, Esbensen K, Geladi P.
\newblock {Principal component analysis}.
\newblock Chemometrics and Intelligent Laboratory Systems. 1987;2(1):37--52.

\bibitem{Kruskal1964}
Kruskal JB.
\newblock {Nonmetric multidimensional scaling: A numerical method}.
\newblock Psychometrika. 1964;29(2):115--129.

\bibitem{tene+00science}
Tenenbaum JB, de~Silva V, Langford JC.
\newblock {A global geometric framework for nonlinear dimensionality
  reduction.}
\newblock Science (New York, NY). 2000;290(5500):2319--23.

\bibitem{coif+05pnas}
Coifman RR, Lafon S, Lee aB, Maggioni M, Nadler B, Warner F, et~al.
\newblock {Geometric diffusions as a tool for harmonic analysis and structure
  definition of data: diffusion maps.}
\newblock Proceedings of the National Academy of Sciences of the United States
  of America. 2005;102(21):7426--31.

\bibitem{scho+98nc}
Sch{\"{o}}lkopf B, Smola A, M{\"{u}}ller KR.
\newblock {Nonlinear Component Analysis as a Kernel Eigenvalue Problem}.
\newblock Neural Computation. 1998;10(5):1299--1319.

\bibitem{Jain:1999:DCR:331499.331504}
{Jain, A} K, {Murty, M} P, {Flynn, P} J.
\newblock {Data clustering: a review}.
\newblock ACM Computing Surveys. 1999;31(3):264--323.

\bibitem{aggarwal2013data}
Aggarwal CC, Reddy CK.
\newblock {Data Clustering: Algorithms and Applications}.
\newblock CRC Press; 2013.

\bibitem{WIDM:WIDM53}
Murtagh F, Contreras P.
\newblock {Algorithms for hierarchical clustering: an overview}.
\newblock Wiley Interdisciplinary Reviews: Data Mining and Knowledge Discovery.
  2012;2(1):86--97.

\bibitem{kmeans1}
Huang Z.
\newblock {Extensions to the k-means algorithm for clustering large data sets
  with categorical values}.
\newblock Data Mining and Knowledge Discovery. 1998;2(3):283--304.

\bibitem{kmeans2}
Jing L, Ng MK, Huang JZ.
\newblock {An entropy weighting k-means algorithm for subspace clustering of
  high-dimensional sparse data}.
\newblock IEEE Transactions on Knowledge and Data Engineering.
  2007;19(8):1026--1041.

\bibitem{kmeans3}
Su MC, Chou CH.
\newblock {A modified version of the K-means algorithm with a distance based on
  cluster symmetry}.
\newblock IEEE Transactions on Pattern Analysis and Machine Intelligence.
  2001;23(6):674--680.

\bibitem{dbscan}
Ester M, Kriegel HP, Sander J, Xu X.
\newblock {A Density-Based Algorithm for Discovering Clusters in Large Spatial
  Databases with Noise}.
\newblock In: Proceedings of the 2nd International Conference on Knowledge
  Discovery and Data Mining. AAAI Press; 1996. p. 226--231.

\bibitem{optics}
Ankerst M, Breunig MM, Kriegel HP, Sander J.
\newblock {Optics: Ordering points to identify the clustering structure}.
\newblock In: ACM Sigmod Record. ACM Press; 1999. p. 49--60.

\bibitem{outlier1}
Zhao X, Liang J, Cao F.
\newblock {A simple and effective outlier detection algorithm for categorical
  data}.
\newblock International Journal of Machine Learning and Cybernetics.
  2014;5(3):469--477.

\bibitem{outlier2}
Yamanishi K, Takeuchi JI, Williams G, Milne P.
\newblock {On-line unsupervised outlier detection using finite mixtures with
  discounting learning algorithms}.
\newblock Data Mining and Knowledge Discovery. 2004;8(3):275--300.

\bibitem{outlier3}
Petrovskiy MI.
\newblock {Outlier detection algorithms in data mining systems}.
\newblock Programming and Computer Software. 2003;29(4):228--237.

\bibitem{outlier4}
{Angiulli, Fabrizio and Pizzuti} C.
\newblock In: Elomaa T, Mannila H, Toivonen H, editors. {Fast Outlier Detection
  in High Dimensional Spaces}. vol. 2431. Berlin, Heidelberg: Springer Berlin
  Heidelberg; 2002. p. 15--27.

\bibitem{outlier5}
Breunig MM, Kriegel HP, Ng RT, Sander J.
\newblock {LOF: identifying density-based local outliers}.
\newblock ACM SIGMOD Record. 2000;29(2):93--104.

\bibitem{outlier6}
Aggarwal CC, Yu PS, Aggarwal CC, Yu PS.
\newblock {Outlier detection for high dimensional data}.
\newblock Proceedings of the 2001 ACM SIGMOD international conference on
  Management of data - SIGMOD '01. 2001;30(2):37--46.

\bibitem{cpc180_2175}
Blum V, Gehrke R, Hanke F, Havu P, Havu V, Ren X, et~al.
\newblock {Ab initio molecular simulations with numeric atom-centered
  orbitals}.
\newblock Computer Physics Communications. 2009;180(11):2175--2196.

\bibitem{prl77_3865}
Perdew JPJ, Burke K, Ernzerhof M, of~Physics D, {Quantum Theory Group Tulane
  University} NOLJ.
\newblock {Generalized Gradient Approximation Made Simple}.
\newblock Physical Review Letters. 1996;77(18):3865--3868.

\bibitem{prl102_73005}
Tkatchenko A, Scheffler M.
\newblock {Accurate Molecular Van Der Waals Interactions from Ground-State
  Electron Density and Free-Atom Reference Data}.
\newblock Physical Review Letters. 2009;102(7):073005.

\bibitem{prl106_118102}
Tkatchenko A, Rossi M, Blum V, Ireta J, Scheffler M.
\newblock {Unraveling the stability of polypeptide helices: Critical role of
  van der Waals interactions}.
\newblock Physical Review Letters. 2011;106(11):118102.

\bibitem{cej19_11224}
Baldauf C, Pagel K, Warnke S, {Von Helden} G, Koksch B, Blum V, et~al.
\newblock {How cations change peptide structure}.
\newblock Chemistry - A European Journal. 2013;19(34):11224--11234.

\bibitem{C4CP05541A}
Schubert F, Rossi M, Baldauf C, Pagel K, Warnke S, von Helden G, et~al.
\newblock {Exploring the conformational preferences of 20-residue peptides in
  isolation: Ac-Ala19-Lys + H(+)vs. Ac-Lys-Ala19+ H(+) and the current reach of
  DFT}.
\newblock Phys Chem Chem Phys. 2015;17(11):7373--7385.

\bibitem{C4CP05216A}
Schubert F, Pagel K, Rossi M, Warnke S, Salwiczek M, Koksch B, et~al.
\newblock {Native like helices in a specially designed $\beta$ peptide in the
  gas phase}.
\newblock Phys Chem Chem Phys. 2015;17(7):5376--5385.

\bibitem{doi:10.1021/jp412055r}
Rossi M, Chutia S, Scheffler M, Blum V.
\newblock {Validation Challenge of Density-Functional Theory for
  Peptides-Example of Ac-Phe-Ala5-LysH(+).}
\newblock The journal of physical chemistry A. 2014;118(35):7349--59.

\bibitem{0953-8984-27-49-493002}
Baldauf C, Rossi M.
\newblock {Going clean: structure and dynamics of peptides in the gas phase and
  paths to solvation.}
\newblock Journal of physics Condensed matter : an Institute of Physics
  journal. 2015;27(49):493002.

\bibitem{ropo_scirep_2016}
Ropo M, Blum V, Baldauf C.
\newblock {Trends for isolated amino acids and dipeptides: Conformation,
  divalent ion binding, and remarkable similarity of binding to calcium and
  lead}.
\newblock arXiv:160602151 [physics, q-bio]. 2016;.

\bibitem{ramachandran-plot}
Ramachandran GN, Ramakrishnan C, Sasisekharan V.
\newblock {Stereochemistry of polypeptide chain configurations.}
\newblock Journal of molecular biology. 1963;7(1):95--99.

\bibitem{fischer_chemical_2000}
Fischer G.
\newblock {Chemical aspects of peptide bond isomerisation}.
\newblock Chemical Society Reviews. 2000;29(2):119--127.

\bibitem{dugave_cistrans_2003}
Dugave C, Demange L.
\newblock {Cis-trans isomerization of organic molecules and biomolecules:
  Implications and applications}.
\newblock Chemical Reviews. 2003;103(7):2475--2532.

\bibitem{weiss_peptide_1998}
Weiss MS, Jabs A, Hilgenfeld R.
\newblock {Peptide bonds revisited}.
\newblock Nature structural biology. 1998;5(8):676.

\bibitem{Dejcp11}
De S, Ghasemi SA, Willand A, Genovese L, Kanhere D, Goedecker S.
\newblock {The effect of ionization on the global minima of small and medium
  sized silicon and magnesium clusters}.
\newblock Journal of Chemical Physics. 2011;134(12):124302.

\bibitem{ideh_mg}
Heidari I, De S, Ghazi SM, Goedecker S, Kanhere DG.
\newblock {Growth and structural properties of MgN (N = 10-56) clusters:
  Density functional theory study}.
\newblock Journal of Physical Chemistry A. 2011;115(44):12307--12314.

\bibitem{ghazi_Na}
Ghazi SM, De S, Kanhere DG, Goedecker S.
\newblock {Density functional investigations on structural and electronic
  properties of anionic and neutral sodium clusters Na N (N = 40-147):
  comparison with the experimental photoelectron spectra}.
\newblock Journal of Physics: Condensed Matter. 2011;23(40):405303.

\bibitem{pascalB80}
Pochet P, Genovese L, De S, Goedecker S, Caliste D, Ghasemi SA, et~al.
\newblock {Low-energy boron fullerenes: Role of disorder and potential
  synthesis pathways}.
\newblock Physical Review B - Condensed Matter and Materials Physics.
  2011;83(8):81403.

\bibitem{arde+15jctc}
Ardevol A, Tribello GA, Ceriotti M, Parrinello M.
\newblock {Probing the unfolded configurations of a $\beta$-hairpin using
  sketch-map}.
\newblock Journal of Chemical Theory and Computation. 2015;11(3):1086--1093.

\bibitem{dataint1}
Ba{\v{s}}karada S, Koronios A.
\newblock {A Critical Success Factor Framework for Information Quality
  Management}.
\newblock Information Systems Management. 2014;31(4):276--295.

\bibitem{dataint2}
{Van Den Broeck} J, Cunningham SA, Eeckels R, Herbst K.
\newblock {Data cleaning: Detecting, diagnosing, and editing data
  abnormalities}.
\newblock PLoS Medicine. 2005;2(10):0966--0970.

\bibitem{dataint3}
Gevorgyan A, Poolman MG, Fell DA.
\newblock {Detection of stoichiometric inconsistencies in biomolecular models}.
\newblock Bioinformatics. 2008;24(19):2245--2251.

\bibitem{dataint4}
Ferretti L, Colajanni M, Marchetti M.
\newblock {Distributed, concurrent, and independent access to encrypted cloud
  databases}.
\newblock IEEE Transactions on Parallel and Distributed Systems.
  2014;25(2):437--446.

\bibitem{De2011}
De S, Willand A, Amsler M, Pochet P, Genovese L, Oedecker S.
\newblock {Energy landscape of fullerene materials: A comparison of boron to
  boron nitride and carbon}.
\newblock Physical Review Letters. 2011;106(22):225502.

\bibitem{cosmogit}
{Code repositories from the Laboratory of Computational Science and Modelling
  at EPFL}; 2014.
\newblock http://epfl-cosmo.github.io/.

\bibitem{libatoms}
{Libatoms}; 2014.
\newblock http://www.libatoms.org/.

\end{thebibliography}
\end{document}